\documentclass[sigconf, nonacm]{acmart}
\usepackage{algorithm}
\usepackage{algpseudocode} 
\usepackage{amsmath,amsfonts}
\AtBeginDocument{%
  }

\usepackage{cleveref}
\usepackage{multirow}
\usepackage{bbm}
\usepackage{xcolor}

\usepackage[colorinlistoftodos]{todonotes}

\newcommand{\agent}{PRL-PUTS}
\begin{document}

\title[\agent\ in Pinterest Recommender Systems]{A Production-Ready RL Framework for Personalized Utility Tuning with Pareto Sweeping in Pinterest Recommender Systems}

\author{Yichu Zhou}
\authornote{These authors contributed equally to this work.}
\affiliation{
  \institution{Pinterest Inc.}
  \city{Seattle}
  \state{Washington}
  \country{USA}
}
\email{yichuzhou@pinterest.com}

\author{Mehdi Ben Ayed}
\authornotemark[1]
\affiliation{%
  \institution{Pinterest Inc.}
  \city{New York City}
  \state{New York}
  \country{USA}
} 
\email{mbenayed@pinterest.com}

\author{Lin Yang}
\authornotemark[1]
\affiliation{%
  \institution{Pinterest Inc.}
  \city{San Francisco}
  \state{California}
  \country{USA}
}
\email{linyang@pinterest.com}

\author{Jiacong He}
\affiliation{%
  \institution{Pinterest Inc.}
  \city{San Francisco}
  \state{California}
  \country{USA}
}
\email{jiaconghe@pinterest.com}

\author{Andreanne Lemay}
\affiliation{%
  \institution{Pinterest Inc.}
  \city{New York City}
  \state{New York}
  \country{USA}
}
\email{alemay@pinterest.com}

\author{Jiaye Wang}
\affiliation{%
  \institution{Pinterest Inc.}
  \city{New York City}
  \state{New York}
  \country{USA}
}
\email{jiayewang@pinterest.com}

\author{Jaewon Yang}
\affiliation{%
  \institution{Pinterest Inc.}
  \city{Palo Alto}
  \state{CA}
  \country{USA}
}
\email{jaewonyang@pinterest.com}

\author{Josie Zeng}
\affiliation{%
  \institution{Pinterest Inc.}
  \city{San Francisco}
  \state{California}
  \country{USA}
}
\email{qianjosiezeng@pinterest.com}

\author{Dhruvil Deven Badani}
\affiliation{%
  \institution{Pinterest Inc.}
  \city{San Francisco}
  \state{California}
  \country{USA}
}
\email{dbadani@pinterest.com}

\author{Yijie Dylan Wang}
\affiliation{%
  \institution{Pinterest Inc.}
  \city{Palo Alto}
  \state{CA}
  \country{USA}
}
\email{dylanwang@pinterest.com}

\author{Jiajing Xu}
\affiliation{%
  \institution{Pinterest Inc.}
  \city{San Francisco}
  \state{California}
  \country{USA}
}
\email{jiajing@pinterest.com}

\author{Charles Rosenberg}
\affiliation{%
  \institution{Pinterest Inc.}
  \city{San Francisco}
  \state{California}
  \country{USA}
}
\email{crosenberg@pinterest.com}

\renewcommand{\shortauthors}{Yichu Zhou et al.}

\begin{abstract}
Large-scale recommenders encode multi-objective trade-offs by combining multiple predicted outcomes into a single utility score. Although this utility layer can be updated independently of the ranker, weight tuning remains largely manual, globally applied, slow to adapt to changing environments and business needs, and hard to govern as priorities shift. We propose \textbf{\agent}, a \textbf{P}roduction-ready, ranker independent \textbf{RL} framework for \textbf{P}ersonalized \textbf{U}tility-weight \textbf{T}uning with Pareto \textbf{S}weeping.  
We cast utility tuning as a one-step, value-based RL problem: given request context, an agent selects a utility-weight vector that re-weights ranker predictions to maximize request-level engagement rewards. 
To visualize performance across the trade-off spectrum and allow decision makers to update the deployed operating policy instantly, we adopt an inference-time Pareto frontier sweeping via a scalarization parameter, producing a family of policies and an empirical Pareto frontier used as a governance artifact for operating policy selection. PRL-PUTS runs in parallel with ranking inference without adding serving latency. We validate PRL-PUTS with offline analysis using unbiased exploration logs and online experiments on Pinterest Homefeed where PRL-PUTS showed significant increases in engagement compared to baseline such as +0.13\% increase in successful session, a core metric for user engagement.
\end{abstract}



\received{20 February 2007}
\received[revised]{12 March 2009}
\received[accepted]{5 June 2009}

\maketitle

\section{Introduction}
Large-scale recommender systems predict multiple objectives simultaneously (e.g., clicks, saves)~\citep{jannach2023survey, zhao2019recommending}, and a downstream \emph{utility layer} aggregates the per-objective predictions into a single score used for ranking. A common implementation is a linear utility: a weighted sum of the predicted outcomes, where the weights encode the desired balance among business goals. This design is attractive because the utility layer is simple, debuggable, and can be updated without retraining the ranker. In practice, this makes the utility layer the primary control surface for responding quickly to shifting priorities, while keeping ranker retraining on a different cycle.

However, the same simplicity creates persistent operational and product debt. Operationally, utility weights are typically selected through ad hoc offline analyses and repeated online experiments, after which decision makers choose among competing metric movements without clear visibility into the set of feasible trade-offs. This process is slow (often taking weeks to months from ideation to launch), hard to reproduce, and difficult to revisit as the environment or priorities change. Technically, the ranker itself is frequently refreshed (e.g. data, features, and calibration), while global utility weights are often treated as long-lived constants; as a result, weights tuned on an earlier distribution can become stale relative to the current ranker and traffic. From a product perspective, global weights are fundamentally non-personalized: they impose the same objective trade-off across heterogeneous user intents and contexts, not because this is optimal, but because manually enumerating and validating context-dependent trade-offs is intractable at production scale.

This paper asks a practical question: \emph{how can we make multi-objective trade-offs fast to update, context-sensitive, and governable while respecting production constraints that limit  model retraining frequency and serving-time complexity?} We propose \textbf{\agent{}}, a \textbf{P}roduction-ready ranker-independent \textbf{RL} framework for \textbf{P}ersonalized \textbf{U}tiliy-weight \textbf{T}uning with Pareto \textbf{S}weeping. \agent\ selects a \emph{utility-weight vector} per request and applies it to the ranker’s existing predictions. We cast this as a one-step, value-based RL problem: given request context, the agent chooses weights to maximize request-level engagement rewards. By restricting actions to utility weights, \agent{} provides a small, reviewable control surface that supports fast operating-policy changes and safe rollback, allowing the ranker and control layer to iterate on different cadences.

Beyond learning a single policy, real world deployments require \emph{governance}: stakeholders must be able to select among achievable trade-offs and update the operating policy as priorities shift~\cite{chen2023controllable}. We therefore adopt \emph{inference-time Pareto frontier sweeping}~\citep{roijers2013survey, hayes2021practical, mossalam2016multi}. \agent{} learns objective-specific value functions for the tasks of interest and exposes a scalarization hyperparameter $\alpha$ at inference time to control their relative importance. Sweeping $\alpha$ offline induces a family of policies from the same trained model and yields an empirical Pareto frontier that stakeholders can use to choose an operating policy aligned with current business priorities.
 
We integrate the proposed control layer into Pinterest Homefeed, where it adjusts only the utility aggregation step and adds no serving latency. We evaluate it offline using unbiased exploration logs from randomized production traffic~\citep{bottou2013counterfactual, zhao2019deep} and validate it with online experiments. We quantify trade-offs between Repin (saves) and P2P  impressions (impressions under related-Pin context) while tracking Successful Sessions, and show strong offline-to-online agreement, with gains attributable to contextual (request-dependent) weight selection rather than a single shifted global weight vector.

This work makes the following contributions:
\begin{itemize}
    
    \item \textbf{Ranker-independent, one-step RL formulation:} We cast utility-weight selection as a one-step, value-based RL problem whose actions are utility-weight vectors applied to ranker predictions, enabling request-time contextual control.

    \item \textbf{Actionable multi-objective governance via Pareto sweeping:} We adopt inference-time sweeping with a scalarization hyperparameter $\alpha$ to induce a family of policies from a single trained model and construct an empirical Pareto frontier for deployment-time operating policy selection.

    \item \textbf{Production integration and deployment:} We present a design that runs alongside ranking inference, adds no serving latency, and is deployed in production on Pinterest Homefeed.

    \item \textbf{End-to-end evidence:} We show that offline Pareto-swept trade-off estimates reliably predict online metric movements, with gains driven by weight personalization rather than shifts in global weights.
\end{itemize}
\section{Related Work}\label{sec:related}

\textit{Utility Weights in Production Ranking.}
Production recommender systems typically aggregate multiple engagement predictions into a single ranking score using a linear utility function with manually tuned weights ~\citep{zhao2019recommending, milli2023choosing}.
In practice, these weights are iterated through offline analysis and A/B tests and then deployed as a largely global, static configuration, which can misspecify trade-offs across heterogeneous users and request contexts~\citep{jeunen2024multi, jannach2023survey, chen2023controllable}. Recent work argues that fixed scalarization enforces a one-size-fits-all trade-off and explores learning context-dependent preferences/dynamic weighting~\citep{yang2025deep, jeunen2024multi,wanigasekara2019learning, cunha2024hybrid}. Most closely related, \citet{yang2025deep} learns a \emph{policy-based} RL controller that selects utility weights for ranking utility tuning in Pinterest ads, illustrating the promise of contextual weight control; in contrast, we use a \emph{value-based} formulation that supports inference-time Pareto sweeping from a single trained model and yields an empirical Pareto frontier for stakeholder-governed operating policy selection.

\textit{RL for Recommender Systems: End-to-end Ranking vs. Control-layer RL.}
RL has been studied extensively for recommendation and ranking, particularly for sequential/session objectives and large decision spaces. Representative examples include RL-to-rank for e-commerce search sessions ~\citep{hu2018reinforcement}, page-wise and whole-chain recommendation formulations ~\citep{zhao2018deep, zhao2020whole}, hierarchical RL for integrated recommendation ~\citep{xie2021hierarchical}, and web-scale in-session optimization frameworks ~\citep{ayed2025recomindreinforcementlearningframework}. Across these lines of work, RL is typically embedded in the core recommendation decision (e.g., selecting items/slates or optimizing multi-step trajectories), which often couples the policy tightly to candidate generation, ranking models, and the broader serving pipeline~\citep{ge2022toward}. In contrast, our RL agent operates as a production-friendly control layer: the action is a utility weight vector applied to existing per-objective predictions, rather than replacing the ranker or choosing items directly. This design reduces coupling with ranking model iteration and better matches stringent serving constraints in web-scale systems, since the policy can be evaluated at request time independently of item-level scoring (consistent with real-time infrastructure considerations discussed in systems work such as ~\citep{liu2022monolith}).

\textit{Multi-objective Control and Pareto-style Operating-policy Selection.}
Multi-objective optimization is inherent in production recommenders, where stakeholders frequently need to trade off competing engagement goals. RL foundations and practical RL surveys emphasize that preference specifications and reward design strongly shape learned behavior, and that changing preferences can be operationally costly if it requires retraining ~\citep{roijers2013survey, sutton2018reinforcement, zhao2019deep}. We focus on enabling post hoc trade-off selection: we learn objective-specific value estimates and then expose a family of trade-off policies at inference time by sweeping a preference parameter, constructing an empirical frontier from which stakeholders can choose an operating policy~\citep{roijers2013survey,hayes2021practical,mossalam2016multi}.

\textit{Production Learning Loop: Exploration Logging and Offline Evaluation.}
A core challenge in production control is that the “correct” action (here, utility weights) is unlabeled; outcomes are only observed under the deployed action, motivating careful exploration and evaluation from logged data. Counterfactual learning provides a standard framing for this setting ~\citep{bottou2013counterfactual}, and off-policy RL emphasizes learning/evaluating from data generated by a different behavior policy ~\citep{Degris2012}. In our setup, we collect logs via constrained randomized exploration over a discrete set of utility-weight actions, and screen policies offline using Reward@HIT, an action-matching (rejection-style) estimator that evaluates only requests where the policy action matches the logged action~\citep{li2011unbiased, li2012unbiased,dudik2012sample}. Industrial RL for ads/allocation provides precedents for constrained exploration and learning from interaction feedback (e.g., bidding/exposure control ~\citep{cai2017real, jin2018real, wu2018budget, zhao2018deepbidding} and pacing/exposure approaches ~\citep{wang2018learning, wei2023rltp, zhao2021dear}), reinforcing the same production pattern: limited exploration, offline screening, online A/B confirmation~\citep{garcia2015comprehensive, brunke2022safe, wachi2024survey}.
\section{Problem Formulation}\label{sec:problem_formulation}

In this section, we formalize ranker-independent utility-weight tuning as a one-step, multi-objective decision problem at the request level. We first define the multi-task ranker and the utility aggregation used for ranking, then specify the one-step RL tuple $(s,a,r)$ together with the state, action, and reward definitions.

\subsection{Utility Tuning for Ranking}

Modern recommender systems commonly use multi-task ranking models that estimate the likelihood of different user engagement outcomes (e.g., clicks, saves). At serving time, these per-objective predictions are aggregated into a single utility score, typically via a weighted linear combination of the model head scores.

\paragraph{Request-level ranking.}
For each request $s$, the system retrieves a candidate set (e.g., $\approx 2{,}000$ items). A multi-task ranker $h$ then produces, for every candidate item $x$, a vector of predicted scores over $m$ objectives (one score per prediction head).
A utility-weight vector $w\in\mathbb{R}^m$ defines a per-item utility score $u$:

\begin{equation}
u(s,x;w)=\sum\limits_{i=1}^m w_i h_i(s,x)
\label{eq:utility_score}
\end{equation}

where $h_i(s,x)$ is the predicted i-th objective from the ranker $h$ based on the request $s$ and item $x$.
The system serves the top-$k$ ranked list based on $u(s,x;w)$. 



In many production systems, utility weights $w$ are selected through offline analysis and repeated online experiments, and are typically applied globally (one weight for all requests) for extended periods. We recast this manual configuration process as a request-level decision problem that selects utility weights, enabling personalization and faster updates as environments and priorities change.

\subsection{One-step RL / Contextual Bandit Formulation}\label{sec:bandit_formulation}
We model utility-weight tuning as a one-step MDP at the request level. Each logged interaction is a tuple$(s, a, r)$, where $s$ is the request context, $a$ is the selected utility-weight action, and $\mathbf{r}\in\mathbb{R}^M$ is a vector of request-level rewards. Since  $a$ affects only the current request and we optimize immediate engagement, we set $\gamma=0$ and do not model state transitions. To enable learning and offline evaluation from logged production data, we collect exploration traffic under a random policy and record action propensities. 

\subsection{State Representation}\label{sec:state_space}
We define the state at the granularity of a single user request. For each request, $s$ comprises serving-time features summarizing the user and request context. Specifically,  $s$ includes:

\begin{enumerate}
    \item \textbf{User information.} Profile signals and long-term preference summaries (e.g., User embeddings).
    \item \textbf{User action history.}  A sequence of the last $N$ user actions, including embeddings of the engaged items and engagement timestamps/action types.
    \item \textbf{Context information.} Serving-time context features (e.g., device type, surface, and request time).
\end{enumerate}
We restrict $s$ to features available at serving time to ensure the learned policy is deployable and that training/evaluation match online inference. We study sensitivity to the state feature set via ablations in Appendix~\Cref{sec:feature_ablation}.

\subsection{Action Space}\label{sec:action_space}
Given a request-level state $s$, the agent selects an action $a$ that specifies the utility weights for the prediction heads. The selected weights are applied only in the utility aggregation step (Eq.~\ref{eq:utility_score}); the upstream ranker and its per-item predictions remain unchanged.

\paragraph{Controlling a subset of utility weights.}
Our framework supports tuning an arbitrary subset (or all) utility weights. In practice, jointly varying many weights in a production serving stack increases complexity and confounding: objectives interact through aggregation and downstream post-processing, making effects harder to attribute and govern. As an initial deployment step, we therefore tune only the two utility heads with the largest contribution to the production utility score, while fixing all other weights to their production values. This reduces action dimensionality, improves interpretability of learned trade-offs, and enables incremental rollout. A breakdown of per-head utility contributions is provided in Appendix~\Cref{sec:head_dist}. The two controlled objectives are \emph{Repin}, the number of Pins a user saves to a board, and \emph{P2P impressions}, the number of impression events in which a user is shown a Pin in a related-Pin context.

\paragraph{Discretizing the action values.}
There is no ground-truth ``correct'' setting of utility weights; their effects are only observable through online outcomes under a complex serving pipeline. We therefore restrict actions to a compact discrete set to (i) bound exploration risk, (ii) make the control surface reviewable, and (iii) support stable offline policy evaluation from logs. To select candidate weight values, we analyze the distribution of each controlled head’s \emph{contribution} to the overall utility score under the production configuration. This informs weight ranges that cover practically meaningful regimes—from near-negligible influence to strong emphasis—while remaining within operationally safe bounds (Appendix~\Cref{sec:action_values}). We then discretize each range by selecting $K$ linearly spaced values.
Finally, we collect training data by running a randomized exploration policy over the resulting discrete action space.

Let $\mathcal{W}^{\text{repin}}$ and $\mathcal{W}^{\text{p2p}}$ denote the discrete candidate sets for the Repin and P2P impression weights, respectively. The action space is:
\begin{equation}
\mathcal{A} = \{(w^{\text{repin}}, w^{\text{p2p}})\mid w^{\text{repin}}\in\mathcal{W}^{\text{repin}},\; w^{\text{p2p}}\in\mathcal{W}^{\text{p2p}}\},
\end{equation}
where $\vert\mathcal{W}^{\text{repin}}\vert=\vert\mathcal{W}^{\text{p2p}}\vert=K$. In this work, we set $K=7$.

\vspace{-0.2cm}
\subsection{Multi-objective Rewards}\label{sec:rewards}
We use vector-valued rewards $\mathbf{r}\in\mathbb{R}^M$. In our Homefeed instantiation, we align rewards with the two controlled objectives and set  $\mathbf{r}=[r^{repin}, r^{p2p}]$. We learn objective-specific value estimates and expose a scalarization parameter $\alpha$ at inference time to induce a \emph{family} of policies (Section~\ref{sec:pareto}); evaluating this family yields an empirical Pareto frontier that stakeholders can use for operating-policy selection.

Let $n^{repin}$ and $n^{p2p}$ denote the resulting request-level counts of Repin and P2P impression events, respectively. To reduce the influence of rare heavy-tail events and stabilize offline estimation and learning, we use clipped binary rewards (Appendix~\Cref{sec:engagement_analysis}):
\begin{equation}
\begin{aligned}
\mathbf{r} &= [r^{repin}, r^{p2p}] \\
r^{repin} &= \min(n^{repin}, 1) \\
r^{p2p} &= \min(n^{p2p}, 1).
\end{aligned}
\end{equation}
\vspace{-0.5cm}
\section{Proposed Methods: \agent}\label{sec:propsoed_method}

In this section, we describe the high-level design of \agent, our inference-time Pareto sweeping procedure for actionable operating-policy selection, and the model architecture used to predict objective-specific values for utility-weight actions.

\subsection{Agent Design (One-step RL / Contextual Bandit)}
\agent{} operates in the one-step decision setting formalized in \Cref{sec:problem_formulation}. For each request context (state) $s\in\mathcal{S}$, the agent selects an action $a\in\mathcal{A}$, where each action corresponds to a \emph{utility-weight vector} $w$ applied on top of a fixed multi-task ranker’s head predictions (Eq.~\ref{eq:utility_score}). The environment returns immediate request-level rewards $\mathbf{r}=[r^{repin}, r^{p2p}]$. Because the action affects only the ranked list shown for the current request and we optimize immediate engagement outcomes, we use a one-step RL formulation where discount factor $\gamma=0$.

In a one-step setting, the objective-specific action-value functions reduce to conditional expected rewards:
\begin{equation}
Q^{repin}(s,a)=\mathbb{E}[r^{repin}\mid s,a],\quad
Q^{p2p}(s,a)=\mathbb{E}[r^{p2p}\mid s,a].
\end{equation}
We therefore learn a $M$-head value function (with $M$=2 in our instantiation)  that predicts these conditional expectations from logged exploration data. This design (i) decouples value estimation across objectives for clearer credit assignment and (ii) together with a restricted action space, exposes a small, reviewable control surface: the policy selects only from pre-defined utility-weight actions, enabling fast and safe iteration.

\textit{Training data support.}
We train \agent{} using unbiased exploration logs collected under a known randomized logging policy $\mu(a\mid s)$ (uniform over $\mathcal{A}$ in our deployment; \Cref{sec:data_collection}). Uniform exploration provides broad action coverage and simplifies offline evaluation, since propensities are constant (\Cref{sec:OfflinePolicyEvaluation}).
\subsection{Inference-time Pareto Sweeping for Governance}\label{sec:pareto}
In production, stakeholders often need to \emph{update multi-objective trade-offs quickly} as priorities shift. In standard multi-objective RL, one fixes a scalarization of the objectives during training, which yields a single policy; changing the trade-off typically requires retraining. \agent{} instead separates learning from selection: we learn objective-specific value functions once, and at serving time choose the operating policy by varying a scalarization parameter $\alpha$, which induces a family of deterministic policies (a Pareto sweep) without retraining. Given a request context $s$ and a scalarization parameter $\alpha\in[0,1]$, we select the utility-weight action that maximizes a linear scalarization of the two predicted values:
\begin{equation}
a^{\star} = \arg\max\limits_{a\in \mathcal{A}}
\left(
\alpha Q_{\theta}^{p2p}(s,a) + (1-\alpha) Q_{\theta}^{repin}(s,a)
\right)
\label{eq:infer}
\end{equation}
Sweeping $\alpha$ over a finite grid $\mathcal{G}\subset[0,1]$ yields a set of candidate policies $\{\pi_\alpha\}_{\alpha\in\mathcal{G}}$ representing different \emph{supported} trade-offs between objectives.

\begin{figure}[] 
    \centering
    \includegraphics[width=0.8\linewidth]{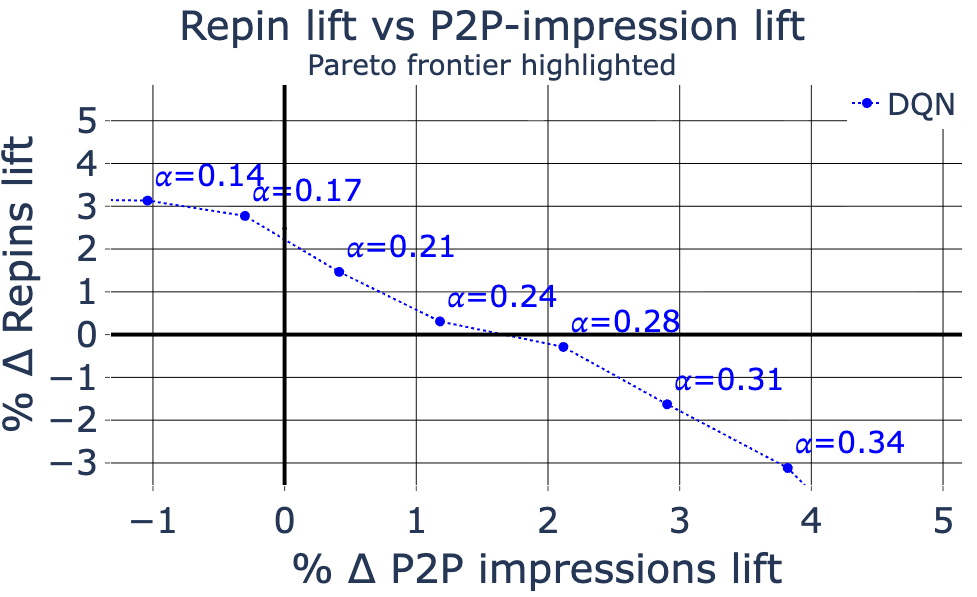}
    \caption{Pareto Frontier. Each data point represents a model with varying Repin and P2P impression lifts compared with production baseline.}
    \label{fig:frontier_example}
\end{figure}

\paragraph{Empirical Pareto frontier construction.}
For governance and deployment planning, we evaluate each $\pi_\alpha$ offline on held-out exploration logs using off-policy evaluation (\Cref{sec:OfflinePolicyEvaluation}). This yields a vector of offline lift estimates $\Delta(\alpha)\in\mathbb{R}^M$ relative to the production baseline, one per objective. We then retain the non-dominated evaluated policies to form an empirical Pareto frontier (\Cref{sec:pareto_frontier_exp}), which stakeholders use to (i) visualize feasible trade-offs (one example is showed in \Cref{fig:frontier_example}) and (ii) select a small set of operating policies for online A/B tests.

\paragraph{Operating-policy control in production.}
Selecting an operating policy requires only setting $\alpha$ via configuration; no model retraining is needed. This makes multi-objective tuning fast to update and easy to roll back.

\begin{algorithm}[t]
\caption{\agent{}: train once, sweep trade-offs, select operating policy}
\label{alg:prl_puts_sweep}
\begin{algorithmic}[1]
\Require exploration logs $D$; action set $\mathcal{A}$; grid $\mathcal{G}\subset[0,1]$; baseline policy $\pi_{\text{prod}}$

\State Train M-head value model $Q_\theta^{repin}(s,a), Q_\theta^{p2p}(s,a)$ on $D$ (\Cref{sec:dqn_arch})
\For{each $\alpha\in\mathcal{G}$}
  \State Define $\pi_\alpha(s)=\arg\max\limits_{a\in\mathcal{A}}\left[\alpha Q^{p2p}_\theta(s,a)+(1-\alpha)Q^{repin}_\theta(s,a)\right]$
  \State Estimate offline lifts of $\pi_\alpha$ vs.\ $\pi_{\text{prod}}$ using OPE (\Cref{sec:OfflinePolicyEvaluation})
\EndFor
\State Construct empirical Pareto frontier; choose $\alpha^\star$ from frontier for online tests/deployment
\end{algorithmic}
\end{algorithm}
Algorithm~\ref{alg:prl_puts_sweep} summarizes our end-to-end training and serving-time control loop for the two-objective instantiation.

\subsection{Two-head Value Model Architecture}\label{sec:dqn_arch}

\begin{figure}[] 
    \centering
    \includegraphics[width=0.8\linewidth]{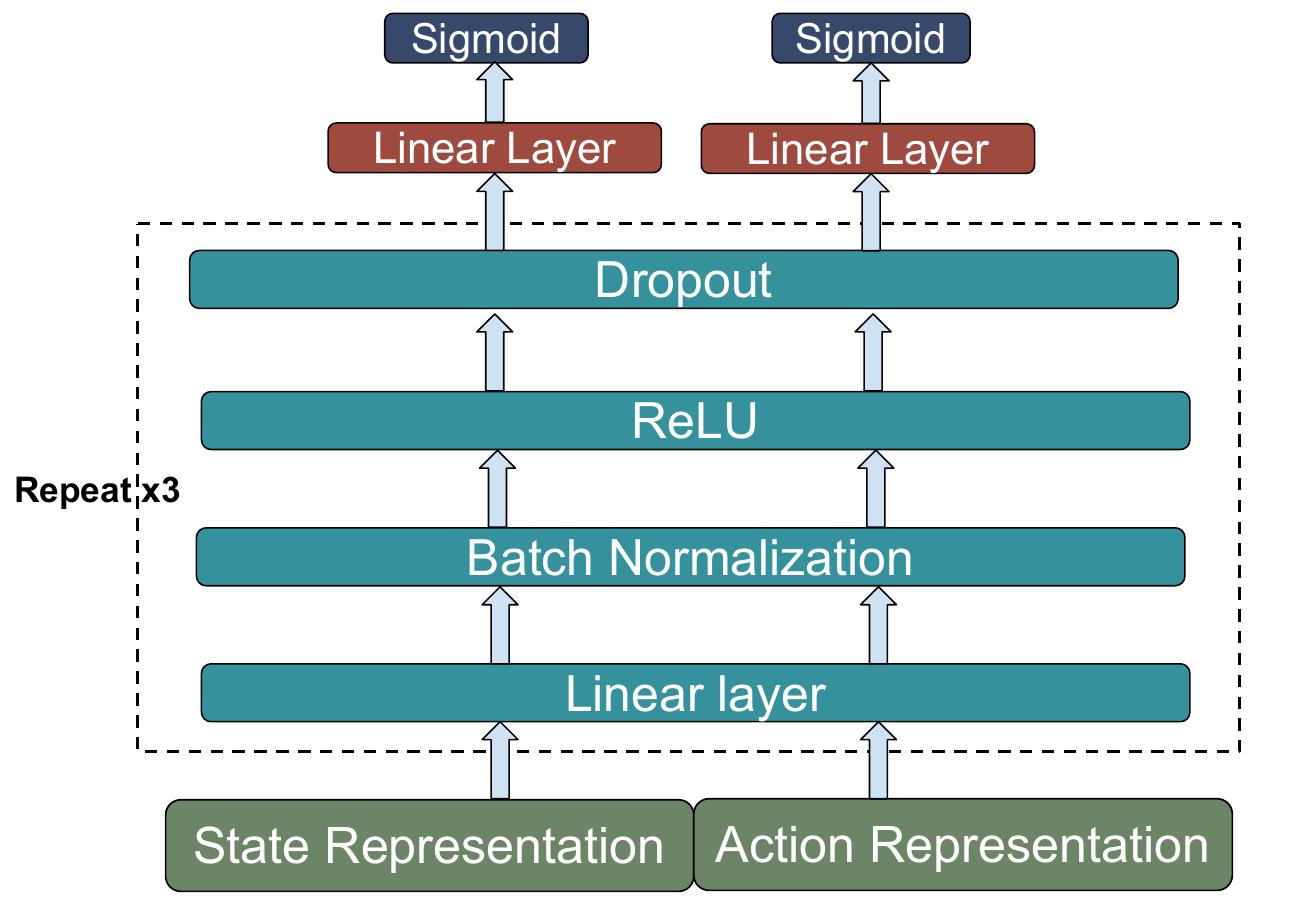}
    \caption{Two Heads Q-Network Architecture. }
    \label{fig:dqn_arch}
\end{figure}

Our model predicts objective-specific values for a given state-action pair $(s,a)$, i.e., estimates of $\mathbb{E}[r^{repin}\mid s,a]$ and $\mathbb{E}[r^{p2p}\mid s,a]$. \Cref{fig:dqn_arch} summarizes the architecture.

\subsubsection{State Module}
The state module constructs a request-level representation from features available at serving time. Categorical features (e.g., device type) are mapped to dense vectors using embedding tables. Sequential features are encoded by a Transformer~\citep{vaswani2017attention}, followed by average pooling to obtain a fixed-length sequence representation. These are concatenated with a user embedding produced by internal Pinterest models and passed through an MLP to produce the final state representation.

\subsubsection{Encoding Actions}\label{sec:encoding_action}
We encode actions as model inputs (rather than enumerating a separate output neuron per action) to keep the design extensible and to facilitate future extensions beyond the current discrete action set. For each action $a=(w^{repin}, w^{p2p})$, we min--max normalize each weight within its candidate set:

\begin{equation}
\begin{aligned}
    w^{repin}_{norm} &= \frac{w^{repin}-\min(\mathcal{W}^{repin})}{\max(\mathcal{W}^{repin})-\min(\mathcal{W}^{repin})}, \\
    w^{p2p}_{norm} &= \frac{w^{p2p}-\min(\mathcal{W}^{p2p})}{\max(\mathcal{W}^{p2p})-\min(\mathcal{W}^{p2p})}
\end{aligned}
\end{equation}

We first apply min--max normalization to the utility weights and then embed the normalized vector $[w^{repin}_{norm}, w^{p2p}_{norm}]$ with a one-layer MLP to obtain a dense action representation. Normalization maps weights to a consistent scale, improving training stability and enabling generalization to other weight values within the same range. At serving time, because $|\mathcal{A}|$ is small (e.g., $K^2$), we can efficiently score all actions and take an argmax under Eq.~\ref{eq:infer}, which keeps inference simple and reliable.

\subsubsection{Backbone Module}
We concatenate the state and action embeddings and feed them into a shared backbone implemented as a three-layer MLP. Each layer consists of batch normalization, a linear projection, and a ReLU nonlinearity, as illustrated in \Cref{fig:dqn_arch}.

\subsubsection{Objective-specific Output Heads}
The model has two output heads sharing the same backbone: one for Repin and one for P2P. Each head is a one-layer MLP producing a scalar value estimate. Since the rewards are clipped to $\{0,1\}$, we bound the predicted values to $[0,1]$ with a sigmoid so they can be interpreted as estimated conditional success probabilities:
\begin{equation}
\begin{aligned}
Q^{repin}_\theta(s,a) &\approx \mathbb{E}\!\left[r^{repin}\mid s,a\right]\in[0,1], \\
Q^{p2p}_\theta(s,a) &\approx \mathbb{E}\!\left[r^{p2p}\mid s,a\right]\in[0,1]
\end{aligned}
\end{equation}
\subsubsection{Training Objective}
Given logged exploration examples $\{(s_i,a_i)\}$, we train the two heads with a sum of per-objective regression losses:

\begin{equation}
L(\theta) = \frac{1}{N}\sum_{i=1}^N\left(
(Q_{\theta}^{p2p}(s_i, a_i)-r_i^{p2p})^2 + (Q_{\theta}^{repin}(s_i, a_i)-r_i^{repin})^2
\right)
\end{equation}

where $\theta$ is the parameters of the model and $r_i^{repin}$ and $r_i^{p2p}$ are the i-th Repin rewards and P2P rewards, $N$ is the total number of examples.
We also experimented with binary cross-entropy (BCE) losses; in our setting BCE did not improve offline or online performance relative to MSE, so we report results with the MSE objective.


\section{Integration into Production}\label{sec:production}

\begin{figure}[] 
    \centering
    \includegraphics[width=1.0\linewidth]{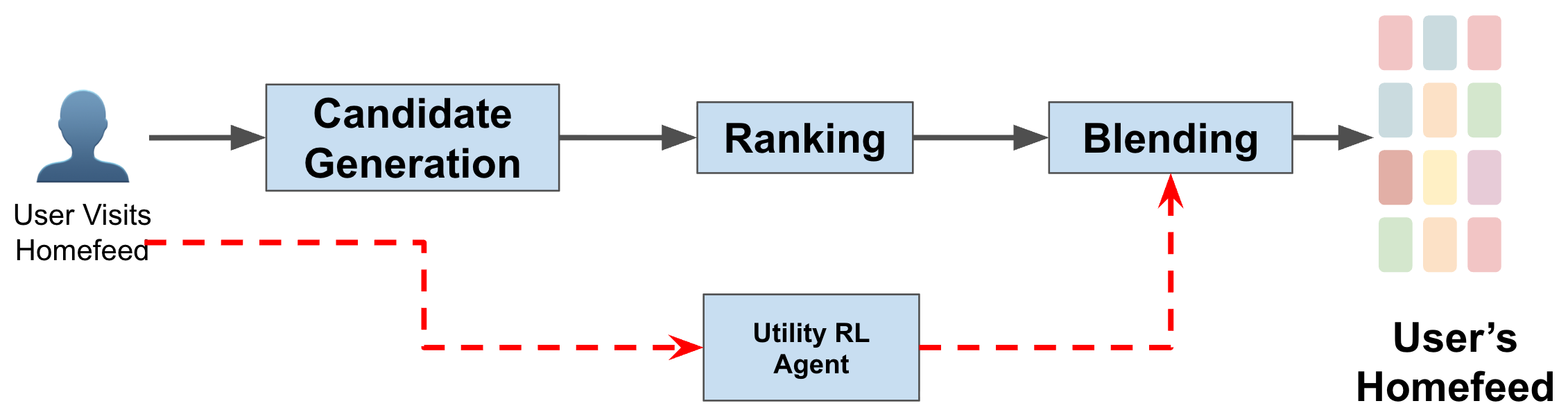}
    \caption{Homefeed Serving Pipeline. The dashed red lines highlight the proposed RL agent pathway, which operates independently of the core ranking process. This decoupled design enables the ranking model and the utility-control agent to be iterated on separately, supports more expressive multi-objective personalization, and introduces negligible additional serving latency.}
    \label{fig:production_structure}
\end{figure}

Our goal is not only to develop an algorithmic approach to utility-weight tuning, but also to demonstrate a practical deployment path in a large-scale production recommender system. In this section, we describe how \agent{} is integrated into Pinterest Homefeed’s serving stack (Figure~\ref{fig:production_structure}) and how the integration enables (i) ranker-independent control, (ii) negligible serving overhead with robust fallback behavior, and (iii) actionable governance via inference-time operating policy selection.
\subsection{Decoupling Ranking Models}\label{sec:upstream_decouple}
A central design goal of \agent{} is to support deployment with minimal disruption to existing ranking infrastructure. We implement \agent{} as a decoupled utility-control layer on top of the existing multi-task ranker: the ranker continues to produce per-objective predictions, while \agent{} selects a request-level utility-weight configuration that is applied only in the downstream utility aggregation step. Both modules run in parallel. This separation makes the framework ranking model-agnostic which enables independent iteration: \agent{} can be enabled, ramped, reconfigured (e.g., changing $\alpha$), or reverted via configuration without changes to the deployed ranker.
\subsection{Serving Latency Overhead}\label{sec:serving_latency}
PRL-PUTS adds no measurable end-to-end serving latency. During serving, the ranker computes per-item predictions while PRL-PUTS runs in parallel to infer a request-level utility-weight vector from request context only. PRL-PUTS does not depend on item-level features or intermediate ranker computations; once per-item predictions are available, the blending layer applies the selected weights to form the final utility score (alongside existing post-processing). For reliability, we enforce a fixed inference budget and fall back to the production static weights on timeout, missing features, or inference failures.
\subsection{Operating Policy Control and Rollback}\label{sec:operating_point}
\agent{} exposes inference-time control through the scalarization parameter $\alpha$ (Section~\ref{sec:pareto}). In production, the deployed operating policy is selected by setting $\alpha$ via serving configuration, allowing stakeholders to switch operating policy (or revert to static weights) without retraining the ranker or the RL model. To make this selection actionable, we operationalize the offline Pareto sweep results as a finite set of candidate operating policies reviewed (e.g., a table and graph over the evaluated grid of $\alpha$ values) with associated offline lift estimates on the primary objectives and key guardrails. We also continuously monitor online key metrics and guardrails (e.g., negative feedback and latency) during ramps and experiments.

\subsection{Ranker refreshes and distribution shift}\label{sec:ranker_refresh}
The integration remains stable under routine ranker iteration as long as the serving contract is preserved: the controlled objective heads retain consistent semantics and remain available at serving time. Because \agent{} operates only on request features and selects among utility-weight configurations, ranker retraining does not require changes in \agent{}. Retraining \agent{} is required only when its contract materially changes (e.g., adding/removing a head controlled by \agent{}) or when substantial distribution shifts in the controlled head predictions invalidate learned context-to-weight mappings. In these cases, the system can safely continue serving the production static utility while collecting fresh exploration data and retraining \agent{}.
\section{Experiments}

\subsection{Data Collection}\label{sec:data_collection}

We log training data from live Pinterest Homefeed traffic with per-request randomization. A capped 1.25\% of requests is used for exploration, where the logging policy $\mu$ samples uniformly from a discrete action set $\mathcal{A}$ ($\mu(a\mid s)=\frac{1}{\vert\mathcal{A}\vert}$) and logs the chosen action and its propensity. Each action $a\in\mathcal{A}$ specifies a utility-weight vector applied to the fixed multi-task ranker’s per-objective predictions. We discretize each tunable weight into $K$ candidate values and form $\mathcal{A}$ as their Cartesian product (all other utility weights are held fixed). For each request, we log the context $s$, action $a$, and request-level rewards for each objective measured from user engagement on the top-$k$ items returned for that request (Repins and P2P impressions).
\subsection{Offline Policy Evaluation}\label{sec:OfflinePolicyEvaluation}
We evaluate candidate policies offline on held-out exploration logs using an off-policy evaluation (OPE) protocol. Exploration traffic samples actions uniformly at random from a discrete action set $\mathcal{A}$ (Section~\ref{sec:data_collection}), so action propensities are constant ($\mu(a\mid s)=1/|\mathcal{A}|$) and require no clipping. We consider deterministic target policies derived from our two-head Q-network via argmax action selection.
\paragraph{Hold-out split.}
We train the Q-network on 14 days of exploration logs and evaluate policies on a disjoint 7-day hold-out period to mitigate temporal leakage and to better reflect generalization under live traffic.





\paragraph{Hit-based OPE / Reward@HIT}
Let the hold-out dataset consist of $N$ logged requests:
$$
D = \{(s_i, a_i, r_i^{p2p}, r_i^{repin})\}_{i=1}^N,
$$
where $s_i$ is the request context, $a_i$ is the logged action, and $r_i^{p2p}$ and $r_i^{repin}$ are the observed request-level rewards. Given a learned Q-network and a scalarization parameter $\alpha$, we derive a deterministic policy $\pi_\alpha$ and its predicted action $a_i^\star=\pi_\alpha(s_i)$ using \Cref{eq:infer}. We define a \emph{hit} event when $a_i^\star=a_i$ and estimate each objective’s expected reward by averaging observed outcomes over hit events:
\begin{equation}
\widehat{V}_{\text{hit}}(\pi_\alpha; r)=
\frac{\sum_{i=1}^{N}\mathbbm{1}[a_i^{\star}=a_i]\cdot r_i}
{\sum_{i=1}^{N}\mathbbm{1}[a_i^{\star}=a_i]},
\label{eq:reward_hit}
\end{equation}
where $r\in\{r^{p2p}, r^{repin}\}$ and $\mathbbm{1}[\cdot]$ is an indicator. Because $\mu(a\mid s)$ is constant under uniform exploration, the inverse-propensity weights cancel in the self-normalized estimator, yielding the hit-based form in Eq.~\ref{eq:reward_hit}.

We choose Reward@HIT for its simplicity and auditability under uniform randomization; we leave more sample-efficient OPE estimators to future work.

\paragraph{Offline lift relative to production.}
For each objective $r$, we report offline lift relative to the production policy $\pi_{\text{prod}}$ (which uses a static weight vector) by evaluating both policies on the same hold-out data:
\begin{equation}
\Delta_{\text{offline}}(\pi_\alpha; r) =
\widehat{V}_{\text{hit}}(\pi_\alpha; r) -
\widehat{V}_{\text{hit}}(\pi_{\text{prod}}; r).
\label{eq:offline_lift}
\end{equation}
This yields paired offline lift estimates for Repin and P2P impression for each evaluated $\alpha$, which we use for model selection and trade-off characterization. Estimator reliability and variance depends on the number of hit events; in our setting, a relatively small action space coupled with the large exploration log and a 7-day hold-out provide sufficient support for stable policy comparisons across $\alpha$.
\subsection{Empirical Pareto Frontier for Governance}\label{sec:pareto_frontier_exp}
Section~\ref{sec:pareto} describes how inference-time sweeping over the scalarization parameter $\alpha$ induces a family of deterministic policies from a single trained two-head Q-network. In this section we describe how we \emph{evaluate} that policy family and convert it into an empirical Pareto frontier that can be used as a governance artifact for deployment decisions.

\paragraph{Discrete sweep and offline evaluation.}
In practice, we evaluate a finite grid of $\alpha$ values, $\mathcal{G}\subset[0,1]$ with $|\mathcal{G}|=25$, and for each $\alpha\in\mathcal{G}$ we derive a deterministic policy $\pi_\alpha$ via argmax action selection using \Cref{eq:infer}. We evaluate each $\pi_\alpha$ on the same 7-day hold-out exploration log using the hit-based OPE estimator in Section ~\ref{sec:OfflinePolicyEvaluation}, yielding paired offline lifts relative to production:
$$
\Delta_{\text{offline}}(\pi_\alpha)=
\big(\Delta_{\text{offline}}(\pi_\alpha; r^{repin}),\,
\Delta_{\text{offline}}(\pi_\alpha; r^{p2p})\big).
$$
Each evaluated $\alpha$ therefore corresponds to one point in the Repin--P2P impression objective space (Repin lift vs.\ P2P impression lift).

\paragraph{Frontier construction.}
We construct an empirical Pareto frontier by retaining only the non-dominated evaluated policies. Specifically, an evaluated policy is dominated if there exists another evaluated policy with at least as large offline lift on both objectives and strictly larger lift on at least one objective. The remaining non-dominated policies form the empirical frontier as shown in \Cref{fig:frontier_example}.

\paragraph{Selecting operating policies for online tests.}
We use the offline frontier to select a small number of representative operating policies spanning different regions of the Repin--P2P impression trade-off curve for online A/B tests. Concretely, we select (i) a Repin-leaning policy near the frontier extreme, (ii) a P2P impression-leaning policy near the opposite extreme, and (iii) a ``knee'' policy that provides a balanced trade-off. To reduce deployment risk and align with stakeholder requirements, we restrict online candidates to operating policies that are predicted offline to be non-degrading on both objectives (i.e., non-negative lift in both Repin and P2P impression).

\subsection{Online Experiments}\label{sec:online_exps}
We validate \agent{} through online experiments in a production setting. The goals are (i) to measure the real-world impact of policies selected from the offline Pareto frontier, and (ii) to test whether offline trade-off estimates reliably predict online metric movements under live traffic.

\subsubsection{Experiment Setup}\label{sec:ab_setup}
We conduct controlled A/B tests by randomly splitting Homefeed traffic at the user level between the production baseline and three treatment policies derived from the same trained two-head Q-network. The baseline uses the current production utility with global static Repin and P2P impression weights. Each treatment reuses the same trained model and serving stack, and differs only in the inference-time policy configuration (i.e., the choice of $\alpha$ selected from the offline Pareto frontier), enabling evaluation of multiple Repin--P2P impression operating policies.

All experiment arms share the same serving infrastructure; the only variation across arms is the action-selection policy applied at serving time. Experiments run for two weeks, with each arm allocated 1\% of total Pinterest Homefeed traffic. We report relative percentage lifts for Repin and P2P impression, and also report Successful Sessions (SS), defined as sessions with at least one key positive action, as a composite engagement metric. Bolded metrics in the tables indicates a statistically significant lift relative to the production baseline.



\subsubsection{Global operating policies from the offline frontier}\label{sec:global_tradeoff}

In the first setting, we deploy a global operating policy where a single trade-off parameter $\alpha$ is applied uniformly across all users receiving a treatment. We select three representative operating policies from the offline Pareto frontier (Repin-leaning, balanced/knee, and P2P impression-leaning). \Cref{tb:global_tradeoff} reports the corresponding offline estimates and online lifts.
\begin{table}[]
\small
\centering
\caption{Global trade-off study (single $\alpha$ for all users). We report relative lifts in Repin, P2P impression, and Successful Sessions (SS) compared with the production baseline.}
\label{tb:global_tradeoff}
\begin{tabular}{lr|rrr}
\toprule
Policy & $\alpha$ & Repin      & P2P  & SS      \\ \midrule
                        &                        & \multicolumn{2}{c}{Online}      \\
Repin-leaning           & 0.17                    & \textbf{+2.26}\%     & -0.21\% & -0.09\%    \\ 
balanced                & 0.21                    & \textbf{+1.35}\%     & -0.04\% & -0.02\%    \\ 
P2P-leaning             & 0.24                    & \textbf{+0.66}\%     & \textbf{+0.30}\% & \textbf{+0.13}\% \\ \midrule
                        &                        & \multicolumn{2}{c}{Offline}     \\
Repin-leaning           & 0.17                    & +2.77\%     & -0.30\%  & -    \\ 
balanced                & 0.21                    & +1.46\%     & +0.41\%  & -  \\ 
P2P-leaning             & 0.24                    & +0.31\%     & +1.18\%  & -  \\ \midrule
                        &                        & \multicolumn{2}{c}{Correlation} \\ 
Correlation             &                        & 0.999          & 0.986            \\ \bottomrule
\end{tabular}
\end{table}

%
As shown in \Cref{tb:global_tradeoff}, online metric movements follow the expected trade-off pattern: Repin-leaning configurations achieve larger Repin lift with some P2P impression reduction, while P2P impression-leaning configurations increase P2P impressions with reduced Repin lift. The balanced operating policy provides a more even trade-off. Across these operating policies, SS is non-degrading and shows a small improvement for the P2P impression-leaning policy.



\subsubsection{Cohort-conditioned operating policies}\label{sec:cohort_tradeoff}
In the second setting, we allow $\alpha$ to vary as a function of a coarse user-context signal (user cohort), enabling different cohorts to adopt different operating policies while remaining interpretable and easy to govern. We define three cohorts:
\begin{itemize}
\item \textbf{CORE:} users who saved on at least 4 days out of the last 28 days.
\item \textbf{CASUAL:} users who were active on at least 4 days out of the last 28 days.
\item \textbf{REST:} all remaining users.
\end{itemize}

\begin{table*}[]
\caption{User cohort study (cohort-specific $\alpha$). We report relative lifts in Repin, P2P impression, and Successful Sessions (SS) compared with the production baseline.}
\label{tb:user_cohorts_exp}
\footnotesize
\begin{tabular}{lrrr|rrrr|rrrr|rrrr}
\toprule
Policy & \multicolumn{3}{c}{$\alpha$} & \multicolumn{4}{c}{Repin}             & \multicolumn{4}{c}{P2P impression}             & \multicolumn{4}{c}{Successful Session (SS)}                \\ \midrule
                        & CORE   & CAUSAL  & REST   & CORE    & CAUSAL  & REST    & TOTAL   & CORE    & CAUSAL & REST   & TOTAL   & CORE    & CAUSAL  & REST    & TOTAL   \\\midrule
                        &        &         &        & \multicolumn{12}{c}{Online}                                                                                         \\ 
repin-leaning           & 0.28  & 0.07   & 0.17  & -0.09\% & \textbf{+7.62\%}  & +1.70\%  & \textbf{+0.65\%}  & -0.30\% & +0.10\% & +0.13\% & -0.03\% & +0.02\%  & -0.03\% & -0.11\% & -0.01\% \\
balanced                & 0.28  & 0.14   & 0.31  & +0.36\%  & \textbf{+3.36\%}  & \textbf{-1.42\%} & +0.36\%  & -0.20\% & \textbf{+0.58\%} & \textbf{+0.55\%} & \textbf{+0.28\%}  & -0.01\% & +0.09\%  & -0.09\% & +0.02\%  \\
p2p-leaning             & 0.28  & 0.18   & 0.35  & -0.11\% & \textbf{+1.90\%}  & \textbf{-2.44\%} & \textbf{-2.19\%} & -0.33\% & \textbf{+1.02\%} & \textbf{+0.73\%} & \textbf{+0.48\%}  & +0.04\%  & \textbf{+0.32\%}  & -0.03\% & \textbf{+0.17\%}  \\ \midrule
                        &        &         &        & \multicolumn{12}{c}{Offline}                                                                                        \\ 
repin-leaning           & 0.28  & 0.07   & 0.17  & +0.02\%  & +10.34\% & +4.87\%  & +1.20\%  & +0.75\%  & +0.49\% & +0.50\% & +0.58\%  & -       & -       & -       & -       \\
balanced                & 0.28  & 0.14   & 0.31  & +0.02\%  & +5.04\%  & +3.12\%  & +0.31\%  & +0.75\%  & +2.53\% & +2.23\% & +1.62\%  & -       & -       & -       & -       \\
p2p-leaning             & 0.28  & 0.18   & 0.35  & +0.02\%  & +2.36\%  & +0.70\%  & 0.00\%  & +0.75\%  & +3.26\% & +3.23\% & +2.02\%  & -       & -       & -       & -       \\\midrule
                        &        &         &        & \multicolumn{12}{c}{Correlation}                                                                                    \\
                        &        &         &        & -       & 0.996   & 0.929   & 0.763   & -       & 0.971  & 0.997  & 0.992   & -       & -       & -       & -      \\ \bottomrule
\end{tabular}
\end{table*}
Operationally, for each request we select $\alpha$ based on the user’s cohort category and apply the corresponding policy. \Cref{tb:user_cohorts_exp} summarizes online results (offline frontiers for each cohort are provided in Appendix~\Cref{sec:user_offline_cohorts}). The CASUAL and REST cohorts respond to $\alpha$ as expected, exhibiting controllable shifts along the Repin--P2P impression trade-off curve (e.g., Repin-leaning increases Repin while P2P impression-leaning increases P2P impression). For the CORE cohort, offline evaluation indicates only one operating policy is non-degrading on both objectives; therefore we reuse that same $\alpha$ value across the three variants.

\subsubsection{Offline-to-online consistency}\label{sec:offline_online}
To quantify whether offline trade-off estimates are decision-useful for selecting online operating policies, we compare offline-predicted lifts to online observed lifts across the evaluated three operating policies and compute Pearson correlation for each objective. We find strong offline-to-online consistency (reported in the final rows of \Cref{tb:global_tradeoff} and \Cref{tb:user_cohorts_exp}), indicating that the offline Pareto frontier is actionable for choosing among operating policies prior to online testing.

\begin{table}[]
\small
\centering
\caption{Matched static weighting v.s. personalized policy.}
\label{tb:static_vs_personalization}
\begin{tabular}{lrrr}
\toprule
                       & Repin   & P2P & SS            \\ \midrule
matched static weights & -0.24\% & +0.07\% & +0.02\%         \\ \midrule
\agent         & +0.12\% & \textbf{+0.21\%} & \textbf{+0.11\%}  \\ \bottomrule
\end{tabular}
\end{table}

\subsubsection{Personalization vs.\ global weight tuning}\label{sec:personalization_vs_static}

A natural question is whether observed gains are driven primarily by global utility reweighting or by contextual personalization. To disentangle these effects, we construct a static baseline by setting Repin and P2P impression utility weights to match the traffic-level averages produced by \agent{}. For example, the learned policy yields an average P2P impression weight of 11.83 (vs.\ the production default of 9.1); we therefore increase the static production weight to 11.83. Results are reported in \Cref{tb:static_vs_personalization}. The static-tuned variant is neutral or negative online and does not reproduce \agent{} gains, indicating that improvements are driven by context-dependent weight selection rather than a single globally retuned weight vector.


\section{Discussion}\label{sec:discussion}

\textit{Known limitations.}
\agent\ is intentionally designed around a constrained, production-compatible abstraction, and that abstraction has clear limits. First, the one-step RL (contextual bandit) formulation optimizes immediate, request-level outcomes and does not explicitly capture longer-horizon effects such as future engagement or retention; it is therefore best suited to objectives with fast feedback and short-window attribution. Second, our instantiation focuses on two controlled objectives, which keeps trade-offs interpretable and governable (a single $\alpha$ sweep), but does not directly address cases with three or more goals. Third, we restrict the controller to a compact discrete action set to bound exploration risk and to ensure adequate action support in logged data for stable offline comparison;
this improves safety and evaluation but reduces expressivity and may miss fine-grained operating policies.

\textit{Longer-horizon decision making.}
A natural extension is to move beyond the one-step assumption when business objectives require delayed credit assignment. Doing so would require additional instrumentation (e.g., longer attribution windows and trajectory-style logging) and learning methods that incorporate bootstrapping over future value rather than regressing only immediate rewards. In large-scale recommenders, longer horizons also increase sensitivity to confounding and non-stationarity, so an important direction is pairing longer-horizon learning with production-suitable safeguards such as conservative deployment constraints and periodic validation of offline-to-online consistency.

\textit{Scaling objectives and improving efficiency.}
Future work can broaden the control surface while improving sample efficiency. For more than two objectives, a single scalarization knob is insufficient; promising directions include constrained selection or exposing a small number of auditable preference parameters. On the action side, moving from discrete grids to continuous weights can improve expressivity, but it raises new challenges in both exploration and off-policy evaluation. A practical path is to combine continuous actions with adaptive, policy-based exploration that concentrates data on high-value regions while maintaining coverage/uncertainty controls needed for trustworthy counterfactual validation.
\section{Conclusions and Future Work}\label{sec:conclusions}
We presented PRL-PUTS, a production-ready RL control layer for request-level utility-weight tuning on top of a fixed multi-task ranker. PRL-PUTS enables fast, ranker-independent, and governable multi-objective control by exposing inference-time Pareto sweeping via a scalarization parameter $\alpha$. Deployed in Pinterest Homefeed with negligible serving overhead, PRL-PUTS delivers controllable Repin--P2P trade-offs and demonstrates strong offline-to-online consistency using unbiased exploration logs and online experiments. Future work will explore extending PRL-PUTS to richer action spaces and longer-horizon objectives.


\begin{acks}
We thank David Woo, Mostafa Keikha, Dafang He and Filip Ryzner for their valuable technical feedback and support.
\end{acks}

\bibliographystyle{ACM-Reference-Format}
\bibliography{sample-base}

@String{Computing = "Computing" }

@inproceedings{cai2017real,
  title={Real-time bidding by reinforcement learning in display advertising},
  author={Cai, Han and Ren, Kan and Zhang, Weinan and Malialis, Kleanthis and Wang, Jun and Yu, Yong and Guo, Defeng},
  booktitle={Proceedings of the tenth ACM international conference on web search and data mining},
  pages={661--670},
  year={2017}
}

@inproceedings{wu2018budget,
  title={Budget constrained bidding by model-free reinforcement learning in display advertising},
  author={Wu, Di and Chen, Xiujun and Yang, Xun and Wang, Hao and Tan, Qing and Zhang, Xiaoxun and Xu, Jian and Gai, Kun},
  booktitle={Proceedings of the 27th ACM International Conference on Information and Knowledge Management},
  pages={1443--1451},
  year={2018}
}

@inproceedings{zhao2021dear,
  title={Dear: Deep reinforcement learning for online advertising impression in recommender systems},
  author={Zhao, Xiangyu and Gu, Changsheng and Zhang, Haoshenglun and Yang, Xiwang and Liu, Xiaobing and Tang, Jiliang and Liu, Hui},
  booktitle={Proceedings of the AAAI conference on artificial intelligence},
  volume={35},
  number={1},
  pages={750--758},
  year={2021}
}

@inproceedings{wei2023rltp,
author = {Wei, Penghui and Chen, Yongqiang and Liu, ShaoGuo and Wang, Liang and Zheng, Bo},
title = {RLTP: Reinforcement Learning to Pace for Delayed Impression Modeling in Preloaded Ads},
year = {2023},
publisher = {Association for Computing Machinery},
address = {New York, NY, USA},
booktitle = {Proceedings of the 29th ACM SIGKDD Conference on Knowledge Discovery and Data Mining},
pages = {5204–5214},
publisher = {Association for Computing Machinery},
location = {Long Beach, CA, USA},
}

@article{sutton2018reinforcement,
  title={Reinforcement learning: An introduction},
  author={Sutton, Richard S},
  journal={A Bradford Book},
  year={2018}
}

@inproceedings{Degris2012,
author = {Degris, Thomas and White, Martha and Sutton, Richard S.},
title = {Off-policy actor-critic},
year = {2012},
address = {Madison, WI, USA},
booktitle = {Proceedings of the 29th International Coference on International Conference on Machine Learning},
pages = {179–186},
numpages = {8},
location = {Edinburgh, Scotland},
series = {ICML'12}
}

@inproceedings{jin2018real,
  title={Real-time bidding with multi-agent reinforcement learning in display advertising},
  author={Jin, Junqi and Song, Chengru and Li, Han and Gai, Kun and Wang, Jun and Zhang, Weinan},
  booktitle={Proceedings of the 27th ACM international conference on information and knowledge management},
  pages={2193--2201},
  year={2018}
}

@misc{ayed2025recomindreinforcementlearningframework,
      title={RecoMind: A Reinforcement Learning Framework for Optimizing In-Session User Satisfaction in Recommendation Systems}, 
      author={Mehdi Ben Ayed and Fei Feng and Jay Adams and Vishwakarma Singh and Kritarth Anand and Jiajing Xu},
      year={2025},
      eprint={2508.00201},
      archivePrefix={arXiv},
      primaryClass={cs.LG},
      url={https://arxiv.org/abs/2508.00201}, 
}

@inproceedings{wang2018learning,
author = {Wang, Weixun and Jin, Junqi and Hao, Jianye and Chen, Chunjie and Yu, Chuan and Zhang, Weinan and Wang, Jun and Hao, Xiaotian and Wang, Yixi and Li, Han and Xu, Jian and Gai, Kun},
title = {Learning Adaptive Display Exposure for Real-Time Advertising},
year = {2019},
publisher = {Association for Computing Machinery},
address = {New York, NY, USA},
booktitle = {Proceedings of the 28th ACM International Conference on Information and Knowledge Management},
pages = {2595–2603},
numpages = {9},
location = {Beijing, China},
series = {CIKM '19}
}

@inproceedings{xie2021hierarchical,
  title={Hierarchical reinforcement learning for integrated recommendation},
  author={Xie, Ruobing and Zhang, Shaoliang and Wang, Rui and Xia, Feng and Lin, Leyu},
  booktitle={Proceedings of the AAAI conference on artificial intelligence},
  volume={35},
  number={5},
  pages={4521--4528},
  year={2021}
}

@article{zhao2019deep,
  title={" Deep reinforcement learning for search, recommendation, and online advertising: a survey" by Xiangyu Zhao, Long Xia, Jiliang Tang, and Dawei Yin with Martin Vesely as coordinator},
  author={Zhao, Xiangyu and Xia, Long and Tang, Jiliang and Yin, Dawei},
  journal={ACM sigweb newsletter},
  volume={2019},
  number={Spring},
  pages={1--15},
  year={2019},
  publisher={ACM New York, NY, USA}
}

@inproceedings{zhao2018deep,
  title={Deep reinforcement learning for page-wise recommendations},
  author={Zhao, Xiangyu and Xia, Long and Zhang, Liang and Ding, Zhuoye and Yin, Dawei and Tang, Jiliang},
  booktitle={Proceedings of the 12th ACM conference on recommender systems},
  pages={95--103},
  year={2018}
}

@inproceedings{zhao2020whole,
  title={Whole-Chain Recommendations},
  author={Zhao, Xiangyu and Xia, Long and Zou, Lixin and Liu, Hui and Yin, Dawei and Tang, Jiliang},
  booktitle={Proceedings of the 29th ACM International Conference on Information \& Knowledge Management},
  pages={1883--1891},
  year={2020}
}

@inproceedings{hu2018reinforcement,
  title={Reinforcement learning to rank in e-commerce search engine: Formalization, analysis, and application},
  author={Hu, Yujing and Da, Qing and Zeng, Anxiang and Yu, Yang and Xu, Yinghui},
  booktitle={Proceedings of the 24th ACM SIGKDD international conference on knowledge discovery \& data mining},
  pages={368--377},
  year={2018}
}

@inproceedings{zhao2018deepbidding,
  title={Deep reinforcement learning for sponsored search real-time bidding},
  author={Zhao, Jun and Qiu, Guang and Guan, Ziyu and Zhao, Wei and He, Xiaofei},
  booktitle={Proceedings of the 24th ACM SIGKDD international conference on knowledge discovery \& data mining},
  pages={1021--1030},
  year={2018}
}

@misc{liu2022monolith,
      title={Monolith: Real Time Recommendation System With Collisionless Embedding Table}, 
      author={Zhuoran Liu and Leqi Zou and Xuan Zou and Caihua Wang and Biao Zhang and Da Tang and Bolin Zhu and Yijie Zhu and Peng Wu and Ke Wang and Youlong Cheng},
      year={2022},
      eprint={2209.07663},
      archivePrefix={arXiv},
      primaryClass={cs.IR},
      url={https://arxiv.org/abs/2209.07663}, 
}

@inproceedings{bottou2013counterfactual,
  author  = {L{{\'e}}on Bottou and Jonas Peters and Joaquin Qui{{\~n}}onero-Candela and Denis X. Charles and D. Max Chickering and Elon Portugaly and Dipankar Ray and Patrice Simard and Ed Snelson},
  title   = {Counterfactual Reasoning and Learning Systems: The Example of Computational Advertising},
  journal = {Journal of Machine Learning Research},
  year    = {2013},
  volume  = {14},
  number  = {101},
  pages   = {3207--3260}
}

@inproceedings{zhao2019recommending,
  title={Recommending what video to watch next: a multitask ranking system},
  author={Zhao, Zhe and Hong, Lichan and Wei, Li and Chen, Jilin and Nath, Aniruddh and Andrews, Shawn and Kumthekar, Aditee and Sathiamoorthy, Maheswaran and Yi, Xinyang and Chi, Ed},
  booktitle={Proceedings of the 13th ACM conference on recommender systems},
  pages={43--51},
  year={2019}
}

@article{milli2023choosing,
  title={Choosing the right weights: Balancing value, strategy, and noise in recommender systems},
  author={Milli, Smitha and Pierson, Emma and Garg, Nikhil},
  journal={arXiv preprint arXiv:2305.17428},
  year={2023}
}

@inproceedings{yang2025deep,
  title={Deep Reinforcement Learning for Ranking Utility Tuning in the Ad Recommender System at Pinterest},
  author={Yang, Xiao and Ayed, Mehdi and Zhao, Longyu and Zhou, Fan and Shen, Yuchen and Engle, Abe and Zhuang, Jinfeng and Leng, Ling and Xu, Jiajing and Rosenberg, Charles and others},
  booktitle={Proceedings of the Nineteenth ACM Conference on Recommender Systems},
  pages={945--948},
  year={2025}
}

@inproceedings{jeunen2024multi,
  title={Multi-objective recommendation via multivariate policy learning},
  author={Jeunen, Olivier and Mandav, Jatin and Potapov, Ivan and Agarwal, Nakul and Vaid, Sourabh and Shi, Wenzhe and Ustimenko, Aleksei},
  booktitle={Proceedings of the 18th ACM Conference on Recommender Systems},
  pages={712--721},
  year={2024}
}

@article{jannach2023survey,
  title={A survey on multi-objective recommender systems},
  author={Jannach, Dietmar and Abdollahpouri, Himan},
  journal={Frontiers in big Data},
  volume={6},
  pages={1157899},
  year={2023},
  publisher={Frontiers Media SA}
}

@inproceedings{chen2023controllable,
  title={Controllable multi-objective re-ranking with policy hypernetworks},
  author={Chen, Sirui and Wang, Yuan and Wen, Zijing and Li, Zhiyu and Zhang, Changshuo and Zhang, Xiao and Lin, Quan and Zhu, Cheng and Xu, Jun},
  booktitle={Proceedings of the 29th ACM SIGKDD conference on knowledge discovery and data mining},
  pages={3855--3864},
  year={2023}
}

@inproceedings{wanigasekara2019learning,
  title={Learning Multi-Objective Rewards and User Utility Function in Contextual Bandits for Personalized Ranking.},
  author={Wanigasekara, Nirandika and Liang, Yuxuan and Goh, Siong Thye and Liu, Ye and Williams, Joseph Jay and Rosenblum, David S},
  booktitle={IJCAI},
  volume={19},
  pages={3835--3841},
  year={2019}
}

@article{cunha2024hybrid,
  title={A Hybrid Meta-Learning and Multi-Armed Bandit Approach for Context-Specific Multi-Objective Recommendation Optimization},
  author={Cunha, Tiago and Marchini, Andrea},
  journal={arXiv preprint arXiv:2409.08752},
  year={2024}
}

@article{roijers2013survey,
  title={A survey of multi-objective sequential decision-making},
  author={Roijers, Diederik M and Vamplew, Peter and Whiteson, Shimon and Dazeley, Richard},
  journal={Journal of Artificial Intelligence Research},
  volume={48},
  pages={67--113},
  year={2013}
}

@article{hayes2021practical,
  title={A practical guide to multi-objective reinforcement learning and planning},
  author={Hayes, Conor F and R{\u{a}}dulescu, Roxana and Bargiacchi, Eugenio and K{\"a}llstr{\"o}m, Johan and Macfarlane, Matthew and Reymond, Mathieu and Verstraeten, Timothy and Zintgraf, Luisa M and Dazeley, Richard and Heintz, Fredrik and others},
  journal={arXiv preprint arXiv:2103.09568},
  year={2021}
}

@article{mossalam2016multi,
  title={Multi-objective deep reinforcement learning},
  author={Mossalam, Hossam and Assael, Yannis M and Roijers, Diederik M and Whiteson, Shimon},
  journal={arXiv preprint arXiv:1610.02707},
  year={2016}
}

@inproceedings{li2011unbiased,
  title={Unbiased offline evaluation of contextual-bandit-based news article recommendation algorithms},
  author={Li, Lihong and Chu, Wei and Langford, John and Wang, Xuanhui},
  booktitle={Proceedings of the fourth ACM international conference on Web search and data mining},
  pages={297--306},
  year={2011}
}

@inproceedings{li2012unbiased,
  title={An unbiased offline evaluation of contextual bandit algorithms with generalized linear models},
  author={Li, Lihong and Chu, Wei and Langford, John and Moon, Taesup and Wang, Xuanhui},
  booktitle={Proceedings of the Workshop on On-line Trading of Exploration and Exploitation 2},
  pages={19--36},
  year={2012},
  organization={JMLR Workshop and Conference Proceedings}
}

@article{dudik2012sample,
  title={Sample-efficient nonstationary policy evaluation for contextual bandits},
  author={Dud{\'\i}k, Miroslav and Erhan, Dumitru and Langford, John and Li, Lihong},
  journal={arXiv preprint arXiv:1210.4862},
  year={2012}
}

@article{garcia2015comprehensive,
  title={A comprehensive survey on safe reinforcement learning},
  author={Garc{\i}a, Javier and Fern{\'a}ndez, Fernando},
  journal={Journal of Machine Learning Research},
  volume={16},
  number={1},
  pages={1437--1480},
  year={2015}
}

@article{brunke2022safe,
  title={Safe learning in robotics: From learning-based control to safe reinforcement learning},
  author={Brunke, Lukas and Greeff, Melissa and Hall, Adam W and Yuan, Zhaocong and Zhou, Siqi and Panerati, Jacopo and Schoellig, Angela P},
  journal={Annual Review of Control, Robotics, and Autonomous Systems},
  volume={5},
  number={1},
  pages={411--444},
  year={2022},
  publisher={Annual Reviews}
}

@article{wachi2024survey,
  title={A survey of constraint formulations in safe reinforcement learning},
  author={Wachi, Akifumi and Shen, Xun and Sui, Yanan},
  journal={arXiv preprint arXiv:2402.02025},
  year={2024}
}

@inproceedings{ge2022toward,
  title={Toward pareto efficient fairness-utility trade-off in recommendation through reinforcement learning},
  author={Ge, Yingqiang and Zhao, Xiaoting and Yu, Lucia and Paul, Saurabh and Hu, Diane and Hsieh, Chu-Cheng and Zhang, Yongfeng},
  booktitle={Proceedings of the fifteenth ACM international conference on web search and data mining},
  pages={316--324},
  year={2022}
}

@article{vaswani2017attention,
  title={Attention is all you need},
  author={Vaswani, Ashish and Shazeer, Noam and Parmar, Niki and Uszkoreit, Jakob and Jones, Llion and Gomez, Aidan N and Kaiser, {\L}ukasz and Polosukhin, Illia},
  journal={Advances in neural information processing systems},
  volume={30},
  year={2017}
}

\appendix
\section{Appendix}
\subsection{Head Contribution Analysis}\label{sec:head_dist}
\begin{figure*}[] 
    \centering
    \includegraphics[width=1.0\linewidth]{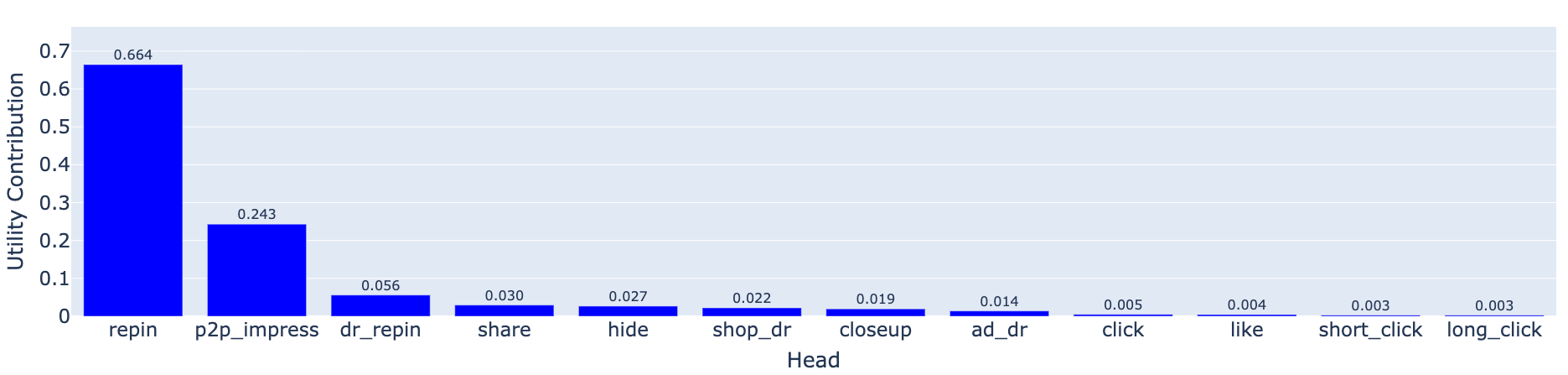}
    \caption{Head contribution distribution. Each bar represents the normalized contribution of each head to the final utility score.}
    \label{fig:head_dist}
\end{figure*}

We compute the utiliy score based on the~\Cref{eq:utility_score}. The contribution of each head ($c_i$) is computed as:

\begin{equation}
c_i(s, x;w) = \frac{\Vert w_ih_i(s,x)\Vert}{\sum\limits_{i}^m\Vert w_ih_i(s,x)\Vert} \\
\end{equation}

where $h_i(s,x)$ is the predicted i-th objective from the ranker $h$ based on the request context $s$ and item $x$.
\Cref{fig:head_dist} shows the contribution of each head using the static prodution weight $w_{prod}$. We observe that Repin and P2P impression head are the top 2 heads that contribute almost $90\%$ to the final utility score. We tune these two heads as the first step.

\subsection{Action Values Selection}\label{sec:action_values}

We select a set of candidate values for the Repin and P2P impression utility weights from the following ranges:
\begin{itemize}
    \item \textbf{Repin}: $[10,200]$
    \item \textbf{P2P impression}: $[1, 30]$
\end{itemize}

\begin{figure*}[] 
    \centering
    \includegraphics[width=1.0\linewidth]{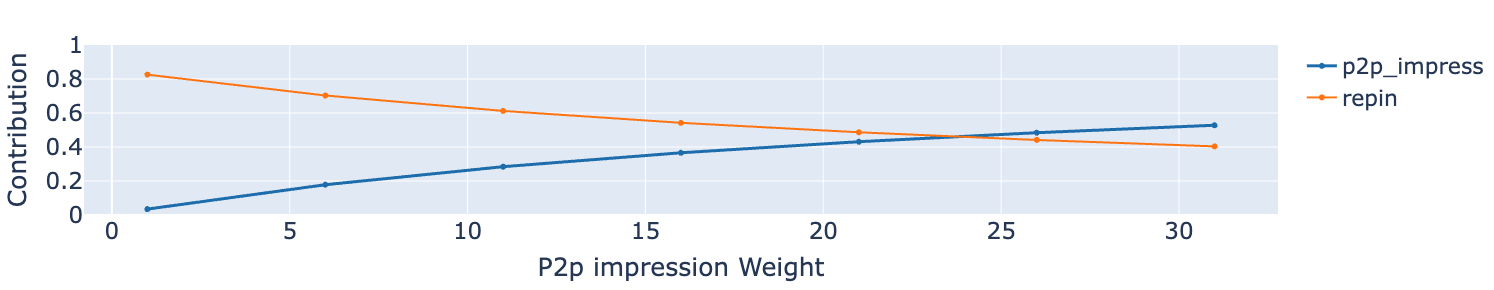}
    \caption{Utility score contribution v.s. P2P impression weight. Vertical axis is the utility score contribution while horizontal axis is the P2P impression weight. We fix the Repin weight to be 91.6. }
    \label{fig:p2p_weight_dist}
\end{figure*}

\begin{figure*}[] 
    \centering
    \includegraphics[width=1.0\linewidth]{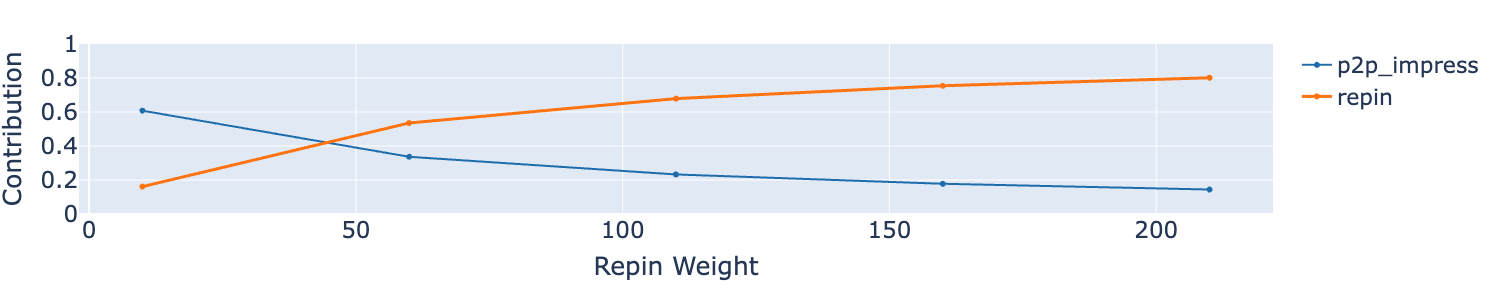}
    \caption{Utility score contribution v.s. Repin weight. Vertical axis is the utility score contribution while horizontal axis is the Repin weight. We fix the P2P impression weight to be 9.1}
    \label{fig:repin_weight_dist}
\end{figure*}

These ranges are chosen to provide broad coverage of the trade-off space. \Cref{fig:repin_weight_dist} and \Cref{fig:p2p_weight_dist} illustrate how each head’s contribution to the overall utility score varies as its corresponding weight changes while holding the other weight fixed. In both cases, the selected ranges span from low to high emphasis, enabling exploration from near-negligible influence to dominant contribution. For example, increasing the P2P impression weight from 1 to 30 changes its contribution from below 1\% to nearly 50\% of the total utility score (see \Cref{fig:p2p_weight_dist}). For reference, the production baseline uses a Repin weight of 91.6 and a P2P impression weight of 9.1.

\subsection{Engagement Distribution Analysis}\label{sec:engagement_analysis}
\begin{figure*}[] 
    \centering
    \includegraphics[width=1.0\linewidth]{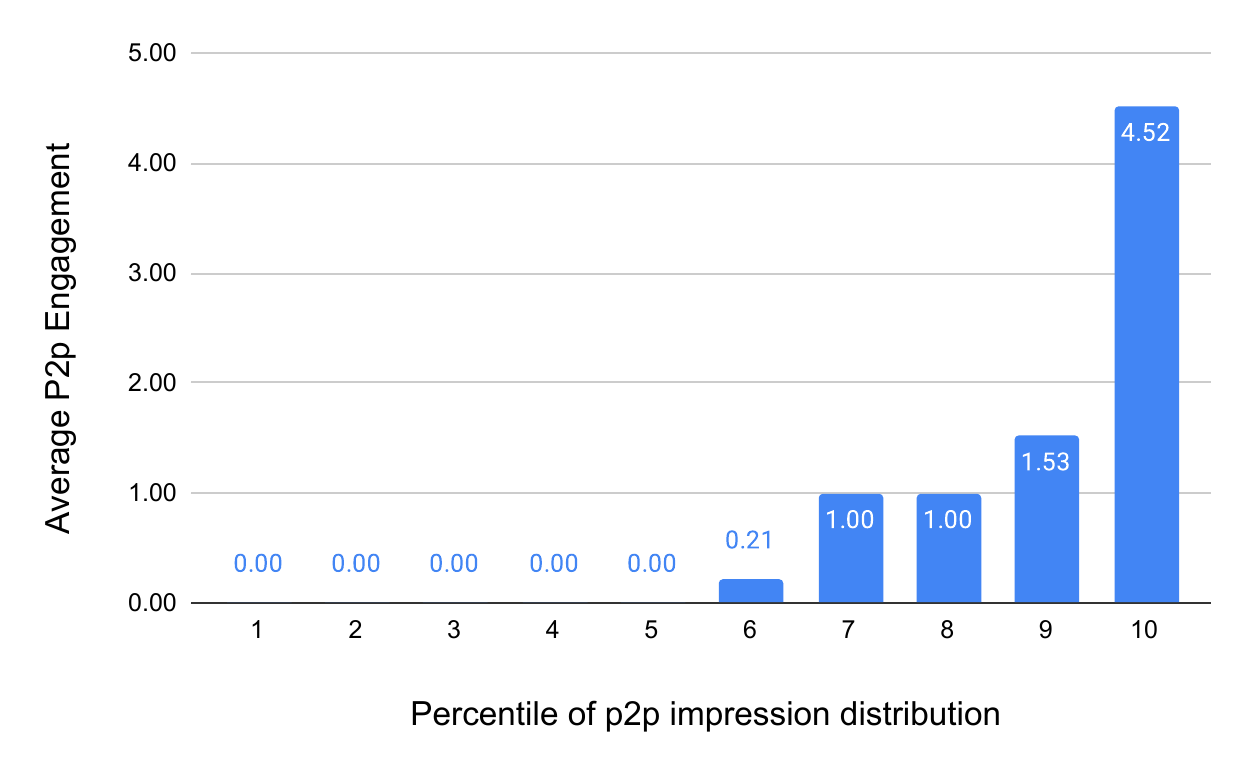}
    \caption{Average P2P impressions within equal-frequency percentile bins of the P2P-impression distribution. Each bar shows the within-bin mean for each decile.}
    \label{fig:p2p_percentile}
\end{figure*}

\begin{figure*}[] 
    \centering
    \includegraphics[width=1.0\linewidth]{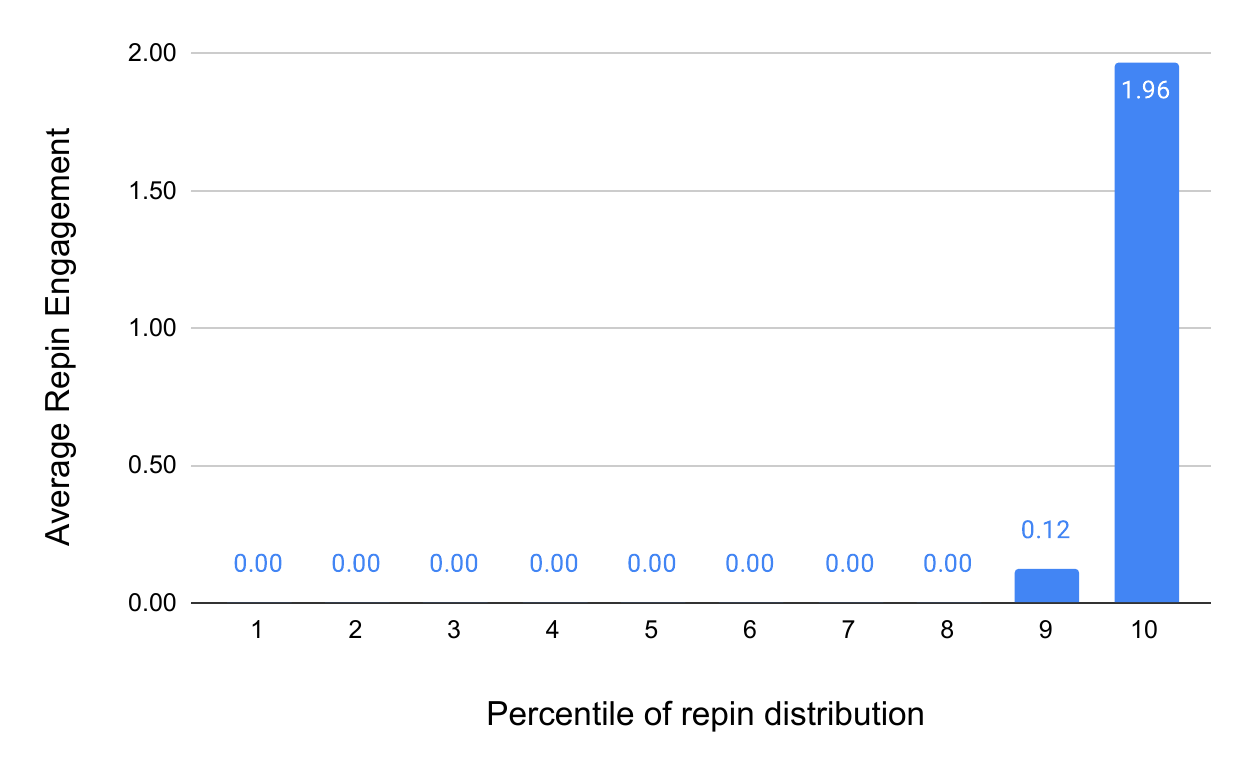}
    \caption{Average Repin within equal-frequency percentile bins of the Repin distribution. Each bar shows the within-bin mean for each decile.}
    \label{fig:repin_percentile}
\end{figure*}

We partition examples into engagement deciles and compute the mean engagement within each decile to summarize how engagement volume is distributed across the population. \Cref{fig:p2p_percentile} and \Cref{fig:repin_percentile} show the percentile distributions of P2P impressions and Repin, respectively. Both metrics exhibit a pronounced long-tailed distribution: more than 80\% of examples receive at most one engagement. To mitigate the influence of extreme values, we define the reward as a clipped version of engagement, mapping counts to the interval $[0,1]$.

This clipping improves robustness but discards information about engagement intensity beyond the first event (e.g., it treats 2 and 200 engagements identically), which may remove useful signal for distinguishing highly engaging content and may limit the agent’s ability to optimize for the upper tail. In future work, we plan to incorporate richer reward shaping that preserves magnitude information—e.g., log-scaled or quantile-normalized rewards, multi-level bins, or separate objectives for occurrence vs. volume—while maintaining training stability.

\subsection{Feature Ablation Study}\label{sec:feature_ablation}

\begin{figure*}[] 
    \centering
    \includegraphics[width=0.8\linewidth]{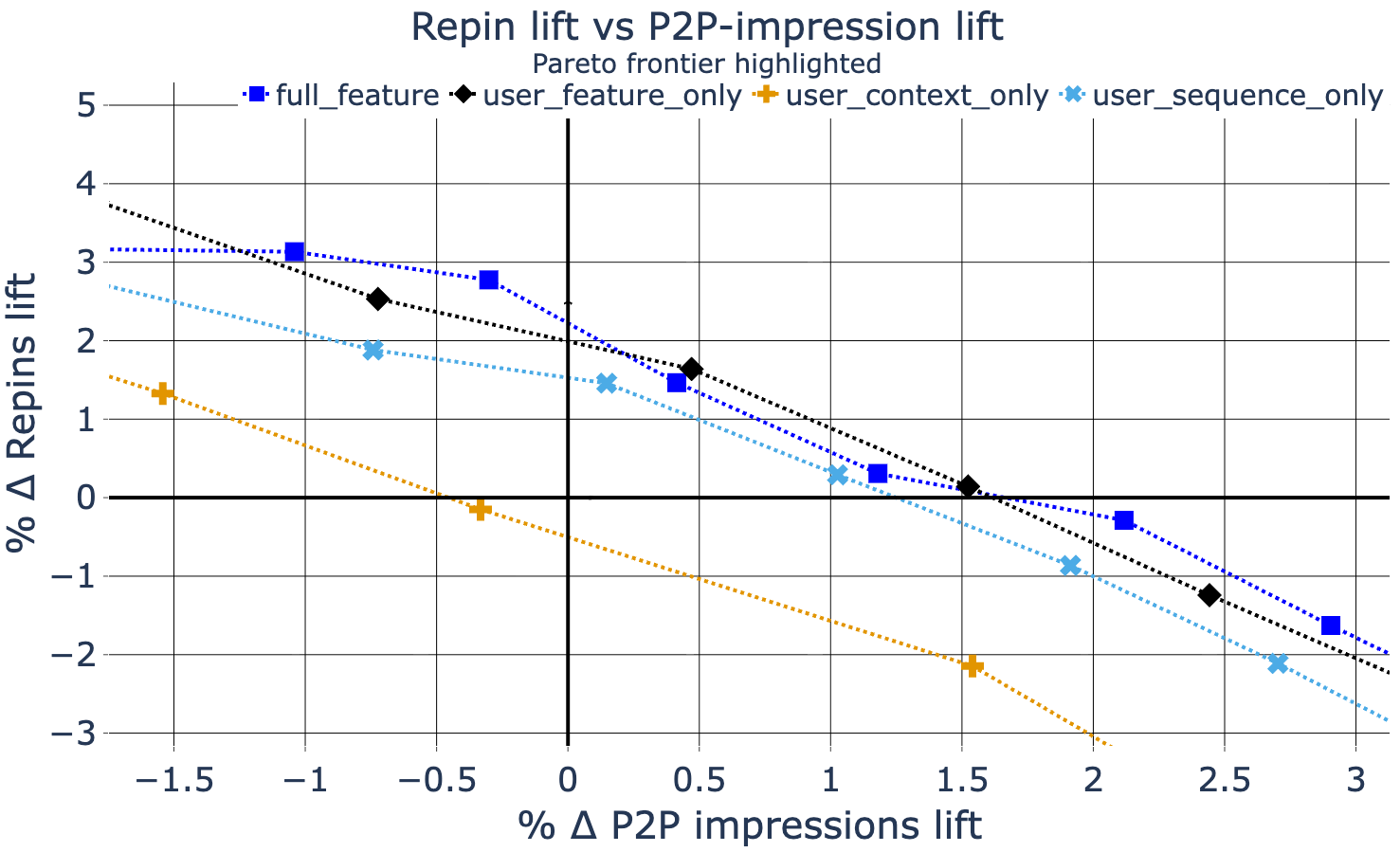}
    \caption{Pareto frontier for feature ablations. Each points represents a different $\alpha$ value applied to the same model.}
    \label{fig:feature_ablation_pareto}
\end{figure*}

We conduct a feature ablation study to quantify the contribution of major feature groups to offline policy performance and to identify which signals are most informative for learning accurate action-value estimates. Unlike a standard leave-one-group-out protocol, our study adopts a \emph{feature-group isolation} setting: starting from a common experimental setup, we train a series of models where each variant is provided with \emph{only one} feature group at a time. This design enables a clearer attribution of performance to individual feature families and highlights the standalone predictive strength of each group under identical training conditions.

\subsubsection{Ablation Setup}
We partition input signals into semantically coherent feature groups (e.g., user features, user action history, and context information). For each ablation variant, we retain exactly one feature group as input and remove all other groups. We keep the remainder of the training and evaluation pipeline unchanged across variants, including data sampling, training schedule, hyperparameters, and model capacity (except for the corresponding input dimensionality change). As a result, differences in offline performance across variants primarily reflect the marginal value of the retained feature group rather than confounding factors introduced by changes in training procedure.

\subsubsection{Ablation Results}

Each ablated variant is evaluated on the same hold-out dataset using Reward@HIT for the policy induced by the corresponding model. For visualization and comparison, we construct and plot the Pareto frontier of each ablation variant alongside the full-feature baseline within a single figure. \Cref{fig:feature_ablation_pareto} shows the results.
When evaluating feature groups in isolation, we find that user information provides the strongest contribution to offline performance. Sequence-based features exhibit trends that are highly consistent with those observed for user features, suggesting they capture complementary signals of similar predictive value. In contrast, models trained using context-only features perform substantially worse, indicating that contextual signals alone are insufficient to support accurate action-value estimation.

\subsection{Offline Pareto Frontier for User Cohorts}\label{sec:user_offline_cohorts}

\begin{figure*}[] 
    \centering
    \includegraphics[width=0.8\linewidth]{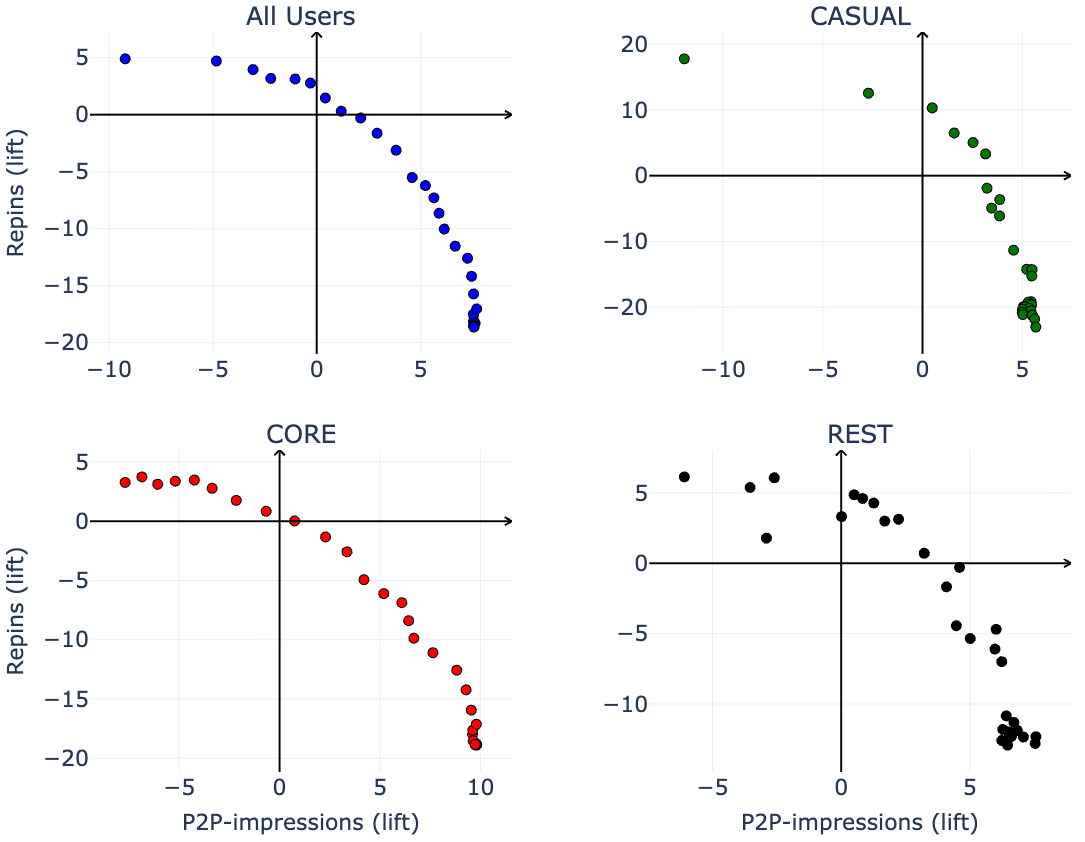}
    \caption{Pareto Frontiers of Repin vs P2P Impression for Each User Cohort}
    \label{fig:user_cohort_offline}
\end{figure*}

\Cref{fig:user_cohort_offline} presents the offline Pareto frontiers for different user cohorts. We observe that the CASUAL and REST cohorts are more sensitive to the trade-off parameter $\alpha$ than the CORE cohort. With appropriate tuning of $\alpha$, we achieve up to a 10\% offline lift in Repin for CASUAL users and up to a 5\% offline lift in Repin for REST users. In contrast, CORE users exhibit a more constrained trade-off: it is difficult to identify operating points that simultaneously improve both Repin and P2P relative to the production baseline. As shown in \Cref{fig:user_cohort_offline}, only a single $\alpha$ value yields a positive lift on both metrics for the CORE cohort.

\end{document}